\documentclass[%
 reprint,
 amsmath,amssymb,
 aps,
]{revtex4-2}

\usepackage{graphicx}
\usepackage{dcolumn}
\usepackage{bm}
\usepackage{hyperref}
\usepackage{xcolor}
\usepackage{tikz}

\begin{document}

\preprint{APS/123-QED}

\title{Collective Alignment in LLM Multi-Agent Systems:\\Disentangling Bias from Cooperation via Statistical Physics}

\author{Cristiano De Nobili}
\altaffiliation[email: ]{cristiano@critiqality.ai}
\affiliation{Critiqality, Milan, Italy}


\begin{abstract}
We investigate the emergent collective dynamics of LLM-based multi-agent systems on a 2D square lattice and present a model-agnostic statistical-physics method to disentangle social conformity from intrinsic bias, compute critical exponents, and probe the collective behavior and possible phase transitions of multi-agent systems.
In our framework, each node of an $L\!\times\!L$ lattice hosts an identical LLM agent holding a binary state ($+1$/$-1$, mapped to \texttt{yes}/\texttt{no}) and updating it by querying the model conditioned on the four nearest-neighbor states.
The sampler temperature~$T$ serves as the sole control parameter.
Across three open-weight models (\texttt{llama3.1:8b}, \texttt{phi4-mini:3.8b}, \texttt{mistral:7b}), we measure magnetization and susceptibility under a global-flip protocol designed to probe $\mathbb{Z}_2$ symmetry.
All models display temperature-driven order-disorder crossovers and susceptibility peaks; finite-size scaling on even-$L$ lattices yields effective exponents $\gamma/\nu$ whose values are model-dependent, close to but incompatible with the 2D Ising universality class ($\gamma/\nu=7/4$). Our method enables the extraction of effective $\beta$-weighted couplings $\tilde{J}(T)$ and fields $\tilde{h}(T)$, which serve as a measure of social conformity and intrinsic bias. In the models we analyzed, we found that collective alignment is dominated by an intrinsic bias ($\tilde{h}\gg\tilde{J}$) rather than by cooperative neighbor coupling, producing field-driven crossovers instead of genuine phase transitions.
These effective parameters vary qualitatively across models, providing compact collective-behavior fingerprints for LLM agents and a quantitative diagnostic for the reliability of multi-agent consensus and collective alignment.
\end{abstract}

\maketitle

\section{Introduction}
\label{sec:intro}

Multi-agent systems (MAS) of large language models are rapidly becoming a foundational architecture for collaborative and collective AI~\cite{guo2024largelanguage,hong2024metagpt}, with applications spanning automated debate~\cite{du2023multiagentdebate,smit2024goingmad}, scientific reasoning~\cite{wang2022selfconsistency}, content moderation~\cite{ruan2025reachingagreementreasoningllm}, collective decision-making~\cite{chuang2024wisdom,piatti2024cooperate}, agents economy~\cite{tomasev2025virtualagenteconomies,wang2026profitredteam}, and decentralized AI~\cite{sun2023cooperativeaidecentralizedcommitment,Ding2025decentralized,riehl2026karma}.
As these systems scale from research prototypes to production deployments, a fundamental question becomes urgent: when a group of LLM agents reaches apparent consensus, is this alignment the result of genuine cooperation, or simply the amplification of shared biases encoded in the underlying model?
The answer has direct consequences for MAS alignment and for the reliability of MAS-based pipelines, where false consensus can propagate errors and suppress diversity of reasoning.

Statistical physics offers a natural and powerful language for these phenomena.
Order parameters quantify the degree of collective alignment, control parameters tune the balance between order and disorder, and universality classes classify macroscopic behavior independently of microscopic details.
The 2D Ising model is the simplest nontrivial system exhibiting a continuous phase transition with well-characterized critical exponents, making it an ideal theoretical reference point.
By placing LLM agents on a square lattice with binary states and nearest-neighbor interactions, we can ask directly whether LLM collectives fall into the Ising universality class or depart from it, and, crucially, \emph{why}.

\textbf{The central contribution of this work is a model-agnostic  statistical-physics method for studying multi-agent systems.}
The method takes as input only the observable update logs of any LLM that can be queried with a structured prompt, requiring no access to the model's internal weights, logits, or architecture.
From these logs alone, the method enables: (i) the disentangling of intrinsic single-agent bias from cooperative inter-agent influence through an effective coupling $\tilde{J}$ and field $\tilde{h}$ extracted via logistic regression; (ii) the computation of critical exponents through standard finite-size scaling of canonical observables such as the magnetization and the susceptibility; and (iii) the broader exploration of statistical and collective behavior, including the diagnosis of phase transitions and crossover phenomena, in arbitrary multi-agent architectures.

In this work, we apply this method to a 2D square-lattice setup across three open-weight models, providing the first systematic finite-size study of LLM agents on a lattice.
We measure canonical observables (magnetization $m(T)$ and susceptibility $\chi(T)$), test finite-size scaling against 2D Ising predictions, and extract $(\tilde{J},\tilde{h})$ trajectories that decompose collective behavior into ``social conformity'' (measured by $\tilde{J}$) and ``intrinsic bias'' (measured by $\tilde{h}$). In the language of recent alignment work, $\tilde{h}$ might be the lattice-scale analog of sycophancy and label bias, while $\tilde{J}$ measures genuine inter-agent influence.
We show that the dominant mechanism behind low-temperature alignment is the effective field, not cooperative coupling, and that the resulting crossover phenomenology is model-dependent and generically non-Ising.

Beyond characterizing collective dynamics, the $(\tilde{J},\tilde{h})$ decomposition serves as a probe of the LLM itself.
Each model traces a distinct trajectory in $(\tilde{J},\tilde{h})$ space as temperature varies, revealing how strongly it follows its neighbors versus its own prior.
These trajectories act as ``collective-behavior fingerprints'' that expose properties of the model (such as the strength of its label bias and its responsiveness to context) that would be difficult to measure through single-agent benchmarks alone.

These results bear directly on multi-agent alignment.
Many MAS architectures (debate, self-consistency, committee voting) implicitly assume that agreement among multiple agents provides stronger evidence for correctness than a single query.
However, when all agents are copies of the same model, they share the same intrinsic biases, encoded in their weights~\cite{santagata2024additionbias,zheng2024calibraeval} and shaped by alignment training such as RLHF~\cite{Christiano2017,Ouyang2022}. A well-documented manifestation of these training-induced biases is sycophancy~\cite{Sharma2023}, the tendency of preference-tuned models to align responses with user-suggested or context-suggested positions; in our framework, this and related label preferences appear as a nonzero effective field.
If these shared biases dominate over genuine inter-agent influence (the $\tilde{h}\gg\tilde{J}$ regime identified in this work), then $N$ agents agreeing carries no more evidential weight than a single agent responding $N$ times: the consensus is correlated through the shared bias, not produced by independent deliberation.
The $(\tilde{J},\tilde{h})$ decomposition provides a quantitative pre-deployment test of whether a given model, prompt, and temperature combination falls in this bias-dominated regime or in the genuinely cooperative regime where collective agreement is informative.
Furthermore, the susceptibility peak identifies the temperature at which the collective is most sensitive to perturbation, a quantity that is directly useful for designing human-in-the-loop MAS protocols: at this temperature, a human overseer can most effectively steer the group's opinion.
Finally, evidence of non-equilibrium dynamics in some models (Sec.~\ref{sec:chi-results} and Appendix~\ref{app:tech}) suggests that standard game-theoretic and statistical-equilibrium frameworks, on which much of alignment theory relies, may not fully capture the behavior of LLM collectives, motivating the development of new theoretical tools at the intersection of non-equilibrium physics and alignment.

Our work connects to several active research directions: LLM-based debate and self-consistency protocols~\cite{wang2022selfconsistency,du2023multiagentdebate,smit2024goingmad}; LLM-as-judge evaluation and its known biases (position, verbosity, self-enhancement)~\cite{LLMJudge,LLMJudgeSurvey}; alignment via RLHF and the systematic preferences it induces, including sycophancy~\cite{Christiano2017,Ouyang2022,Sharma2023}; opinion-dynamics models on networks (voter model, majority rule, bounded confidence)~\cite{castellano2009statistical,mullick2025sociophysics,starnini2025opinion,sastre2015majority,dornic2001voter}, the study of LLM calibration and intrinsic biases~\cite{santagata2024additionbias,zheng2024calibraeval,germani2025sourceframing,nadeem2026biasbeyondborders,knipper2025biasdetailsassessmentcognitive}, simulating opinion dynamics with LLM agents and emergent cultural attractors~\cite{chuang2024wisdom,perez2025telephone,johnson2026increasing}, mean-field treatments of multi-agent learning and behavior modeling~\cite{yang2018mean,cozzi2025learning}, agents economy~\cite{tomasev2025virtualagenteconomies,wang2026profitredteam}, decentralized AI~\cite{sun2023cooperativeaidecentralizedcommitment,Ding2025decentralized,riehl2026karma}, and recent investigations into the equilibrium structure of LLM-driven dynamics~\cite{song2025detailedbalance}.

\section{Model and Methods}
\label{sec:model}

\subsection{Lattice, states, and update rule}

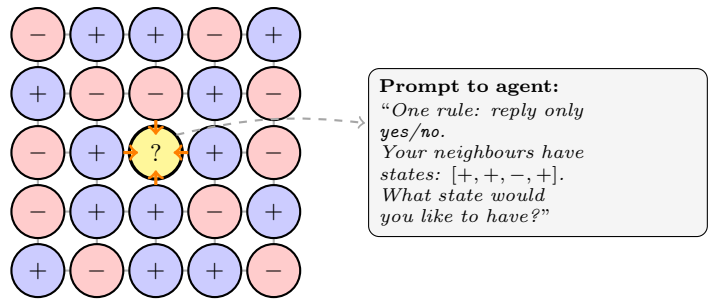
\begin{figure}[t]
\centering
\begin{tikzpicture}[
  scale=0.78,
  spin/.style={circle, draw=black, thick, minimum size=7mm, font=\small\bfseries, inner sep=0pt},
  spinup/.style={spin, fill=blue!20},
  spindown/.style={spin, fill=red!20},
  central/.style={spin, fill=yellow!50, draw=black, very thick},
  bond/.style={-, thick, gray!60}
]

\foreach \x in {0,1,2,3}
  \foreach \y in {0,1,2,3,4}
    \draw[bond] (\x,\y) -- (\x+1,\y);
\foreach \x in {0,1,2,3,4}
  \foreach \y in {0,1,2,3}
    \draw[bond] (\x,\y) -- (\x,\y+1);

\node[spinup] at (0,0) {$+$};
\node[spindown] at (1,0) {$-$};
\node[spinup] at (2,0) {$+$};
\node[spinup] at (3,0) {$+$};
\node[spindown] at (4,0) {$-$};

\node[spindown] at (0,1) {$-$};
\node[spinup] at (1,1) {$+$};
\node[spinup] at (2,1) {$+$};
\node[spindown] at (3,1) {$-$};
\node[spinup] at (4,1) {$+$};

\node[spindown] at (0,2) {$-$};
\node[spinup] at (1,2) {$+$};
\node[central] at (2,2) {$?$};
\node[spinup] at (3,2) {$+$};
\node[spindown] at (4,2) {$-$};

\node[spinup] at (0,3) {$+$};
\node[spindown] at (1,3) {$-$};
\node[spindown] at (2,3) {$-$};
\node[spinup] at (3,3) {$+$};
\node[spindown] at (4,3) {$-$};

\node[spindown] at (0,4) {$-$};
\node[spinup] at (1,4) {$+$};
\node[spinup] at (2,4) {$+$};
\node[spindown] at (3,4) {$-$};
\node[spinup] at (4,4) {$+$};

\draw[->, very thick, orange] (1.45,2) -- (1.7,2);
\draw[->, very thick, orange] (2.55,2) -- (2.3,2);
\draw[->, very thick, orange] (2,1.45) -- (2,1.7);
\draw[->, very thick, orange] (2,2.55) -- (2,2.3);

\node[draw, fill=gray!8, align=left, anchor=west, text width=4.2cm,
      rounded corners, font=\scriptsize, inner sep=4pt] (callout) at (5.6, 2)
  {\textbf{Prompt to agent:}\\[1pt]
   ``\emph{One rule: reply only}\\
   \emph{\texttt{yes}/\texttt{no}.}\\
   \emph{Your neighbours have}\\
   \emph{states: $[+,+,-,+]$.}\\
   \emph{What state would}\\
   \emph{you like to have?}''};

\draw[->, thick, dashed, gray!70] (2.35,2.3) to[bend left=10] (5.55,2.5);

\end{tikzpicture}
\caption{Schematic of the LLM-on-lattice setup. Each site of the $L\times L$ square lattice (periodic boundary conditions) hosts an identical LLM agent in a binary state $s_i\in\{+1,-1\}$, mapped to \texttt{yes}/\texttt{no} (blue for $+$, red for $-$). At each micro-update, a site is chosen at random (yellow, central) and its agent is queried with the prompt shown on the right, conditioned on its four nearest-neighbor states (orange arrows). The reply is parsed and written back as $s_i'$. One \emph{sweep} consists of $N=L^2$ such micro-updates.}
\label{fig:lattice-sketch}
\end{figure}

We consider an $L\times L$ square lattice with periodic boundary conditions, $N=L^2$ sites.
Each site~$i$ holds a binary state $s_i\in\{-1,+1\}$, mapped to textual labels (\texttt{no},\,\texttt{yes}).
Updates follow an asynchronous Monte~Carlo-like protocol: at each micro-step a site $(i,j)$ is chosen uniformly at random and its LLM agent is queried with a minimal prompt of the form
\begin{quote}
\small\texttt{One rule: reply only yes/no.  Your neighbours have states: [\ldots].  What state would you like to have?}
\end{quote}
The reply is parsed to $s_i'\in\{-1,+1\}$ and written back.
One \emph{sweep} consists of $N$ such updates.
The LLM sampler temperature~$T$ is the sole external control parameter.
Figure~\ref{fig:lattice-sketch} provides a schematic illustration of the setup.

For reference, the classical 2D Ising Hamiltonian is
\begin{equation}
H = -J\sum_{\langle i,j\rangle} s_i s_j - h\sum_i s_i\,,
\label{eq:ising-H}
\end{equation}
where $J$ is the nearest-neighbor coupling and $h$ an external field.
Onsager's exact solution at $h=0$ gives a continuous phase transition at $\beta_c J = \tfrac{1}{2}\ln(1+\sqrt{2})\approx 0.4407$, with universal critical exponents $\gamma/\nu = 7/4$ in two dimensions.

\subsection{Global-flip protocol}
\label{sec:flip}

LLMs generically prefer one label over the other (e.g., \texttt{yes} over \texttt{no}), introducing a nonzero effective field that breaks $\mathbb{Z}_2$ symmetry. 
To probe and partially cancel this bias, we run paired ensembles with a global relabeling $s_i\to -s_i$ (equivalently swapping \texttt{yes}$\leftrightarrow$\texttt{no} in all prompts and responses. More in Appendix~\ref{app:flip}). Averaging even moments ($\langle m^2\rangle$, $\langle m^4\rangle$) across both flip sectors cancels odd-in-$m$ bias in expectation, while odd moments and the absolute magnetization $\langle|m|\rangle$ can be compared across sectors to quantify the residual field.

\begin{figure}[t]
\centering
\includegraphics[width=\columnwidth]{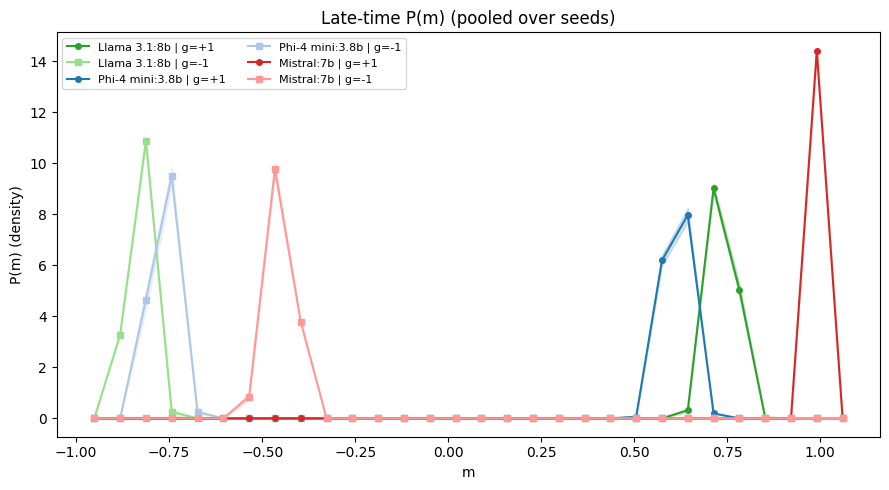}
\caption{Late-time magnetization distribution $P(m)$ pooled over seeds, separated by flip sector ($g=+1$, $g=-1$), for all three models at their respective susceptibility-peak temperatures ($T=0.225$ for \texttt{phi4-mini:3.8b}, $T=0.45$ for \texttt{llama3.1:8b}, $T=1.2$ for \texttt{mistral:7b}; all at $L=20$). \texttt{phi4-mini:3.8b} and \texttt{llama3.1:8b} show approximate mirror symmetry between sectors, validating the global-flip protocol. \texttt{mistral:7b} shows strongly broken symmetry: the $g=+1$ sector locks at $m\approx 1$ while the $g=-1$ sector reaches only $m\approx -0.5$, reflecting its persistent effective field.}
\label{fig:Pm}
\end{figure}

Figure~\ref{fig:Pm} illustrates the effectiveness of this protocol by showing the late-time magnetization distribution $P(m)$ for all three models, separated by flip sector ($g=+1$ and $g=-1$), each measured at the respective susceptibility-peak temperature.
For \texttt{phi4-mini:3.8b} ($T=0.225$) and \texttt{llama3.1:8b} ($T=0.45$), the two sectors produce approximately mirror-symmetric distributions centered at $m\approx \pm 0.65$ and $m\approx \pm 0.8$ respectively, with zero overlap between sectors.
This approximate symmetry validates the pooling procedure for even-moment observables.
For \texttt{mistral:7b} ($T=1.2$), the symmetry is visibly broken: the $g=+1$ sector concentrates at $m\approx +1.0$ (near-complete alignment) while the $g=-1$ sector peaks at $m\approx -0.5$ (partial alignment).
This asymmetry reflects \texttt{mistral:7b}'s persistently strong effective field (Sec.~\ref{sec:Jh}), which prevents the global-flip protocol from fully equalizing the two sectors.

\subsection{Observables}
\label{sec:obs}

From the magnetization time series $m(t)=N^{-1}\sum_i s_i(t)$ after a burn-in window, we estimate late-time averages $\langle m^k\rangle$ by averaging over sweeps and then over seeds.
We compute two susceptibility variants,
\begin{equation}
\chi = \frac{N}{T}\bigl(\langle m^2\rangle - \langle m\rangle^2\bigr),\quad
\chi_{|m|} = \frac{N}{T}\bigl(\langle m^2\rangle - \langle |m|\rangle^2\bigr).
\label{eq:chi}
\end{equation}
When both flip sectors are pooled, $\langle m\rangle\approx 0$ by construction, so $\chi$ reduces to $N\langle m^2\rangle/T$ and diverges trivially as $1/T$ at low temperature.
The absolute susceptibility $\chi_{|m|}$ avoids this artifact by subtracting $\langle|m|\rangle^2$, which correctly captures the magnitude of alignment in both sectors; we therefore use $\chi_{|m|}$ throughout.
In finite-size scaling, a genuine 2D Ising transition predicts $\chi_{\max}(L)\sim L^{\gamma/\nu}$ with $\gamma/\nu=7/4=1.75$.

\begin{figure*}[t]
  \centering
  \includegraphics[width=0.325\textwidth]{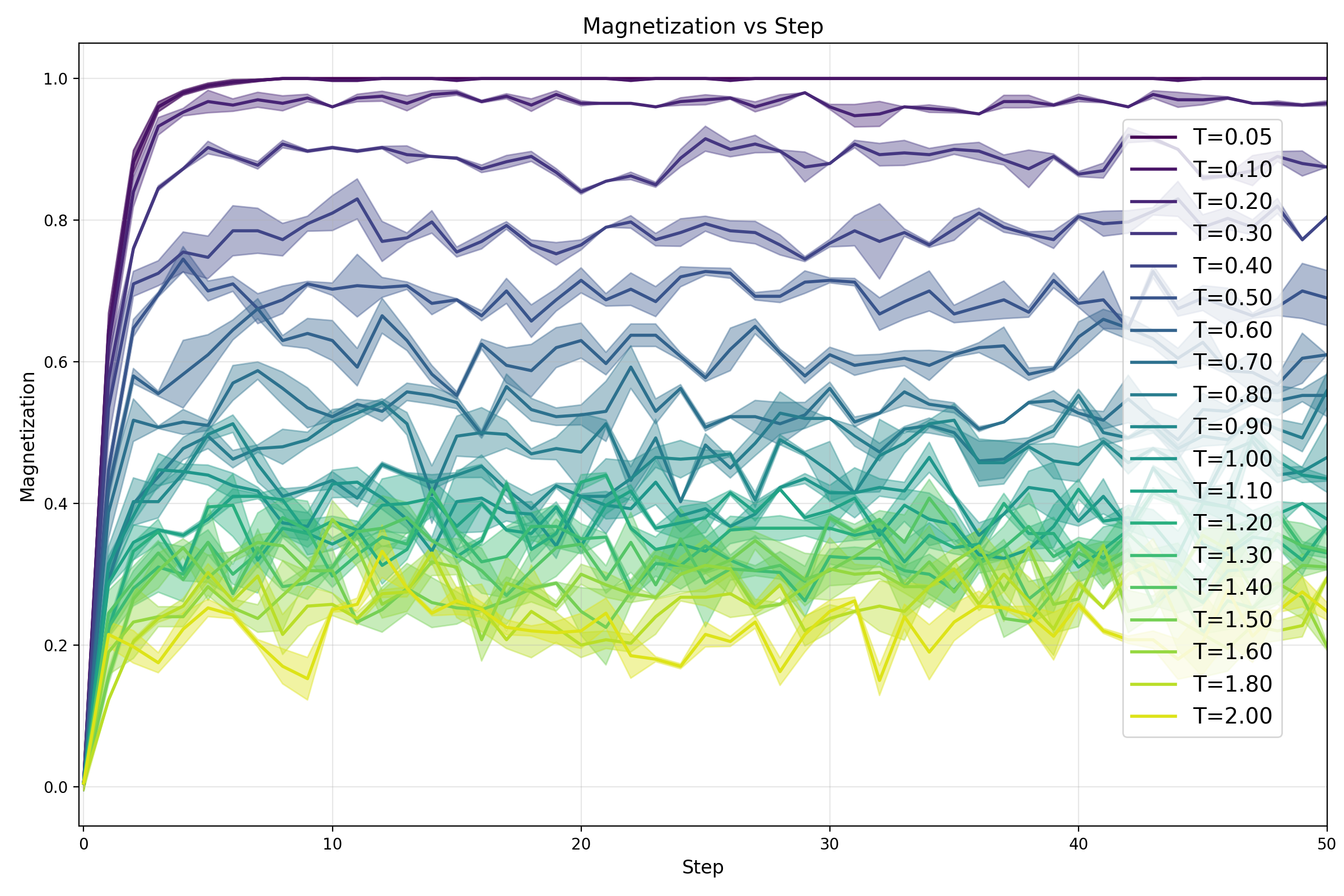}%
  \hfill
  \includegraphics[width=0.325\textwidth]{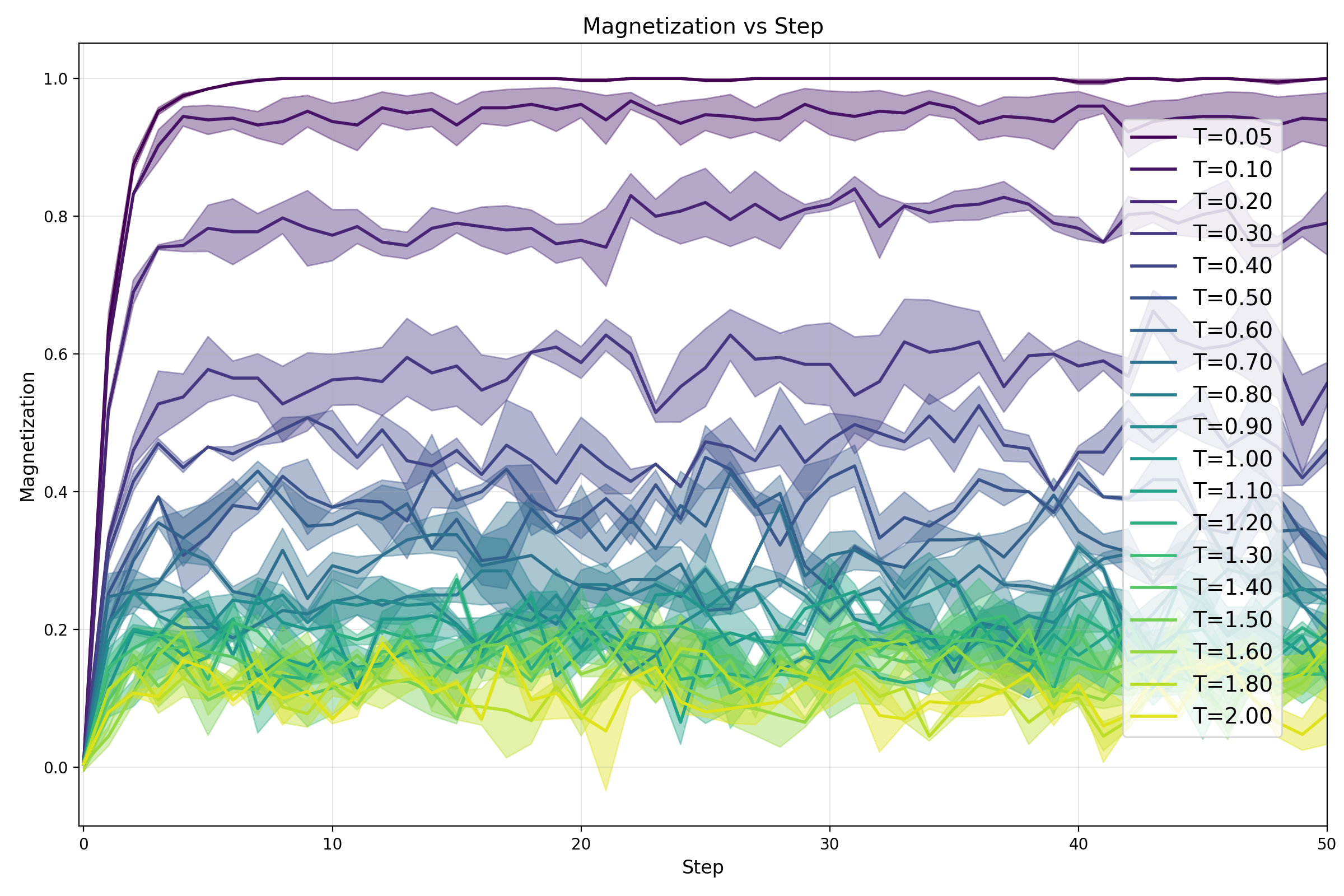}%
  \hfill
  \includegraphics[width=0.325\textwidth]{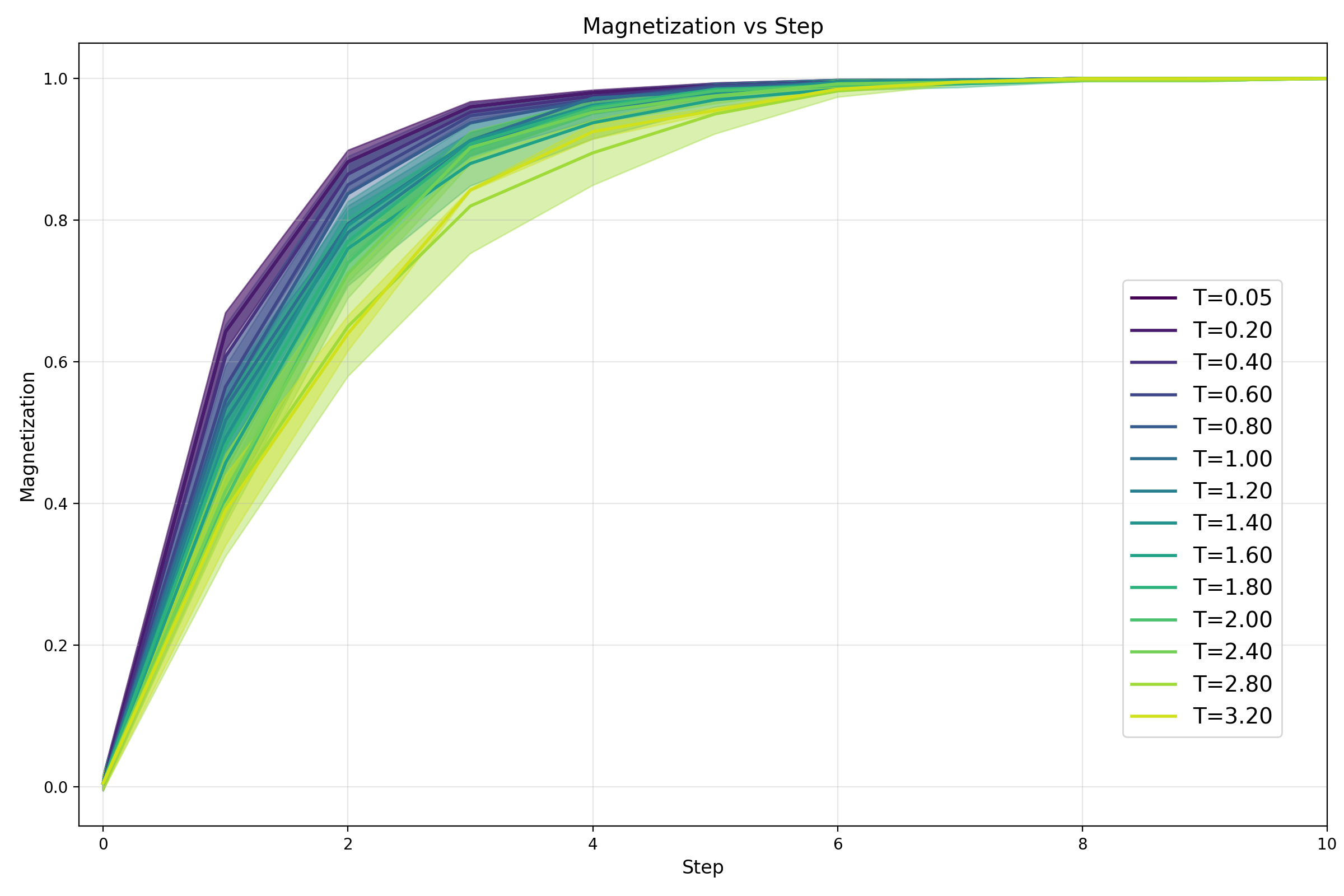}
  \caption{\label{fig:mag-timeseries}%
Magnetization $m(t)$ versus sweep number at various sampler temperatures for ~\texttt{llama3.1:8b} (a), ~\texttt{phi4-mini:3.8b} (b), and ~\texttt{mistral:7b} (c). Shaded bands denote $\pm 1\sigma$ across seeds. \texttt{llama3.1:8b} and \texttt{phi4-mini:3.8b} show clear temperature-driven disordering, while \texttt{mistral:7b} reaches near-full alignment at all temperatures, a direct manifestation of its persistently strong effective field (Fig.~\ref{fig:Jh}).}
\end{figure*}

\subsection{Effective coupling and field from update logs}
\label{sec:kernel}

For each micro-update we record the local neighbor field $k=\sum_{j\in\partial i} s_j \in \{-4,-2,0,2,4\}$ and the resulting spin~$s_i'$.
Aggregating over all updates at a given~$T$, we fit the log-odds to a linear model,
\begin{equation}
\operatorname{logit}\,P(s_i'=+1\mid k) \;\approx\; 2(\tilde{h}+\tilde{J}\,k)\,,
\label{eq:logit}
\end{equation}
where $\tilde{J}=\beta J$ and $\tilde{h}=\beta h$ are effective $\beta$-weighted parameters.
This logistic form is exact for a heat-bath update of the Ising Hamiltonian~\eqref{eq:ising-H}; applied to LLM updates, it provides a model-agnostic decomposition into coupling strength (``social conformity'') and bias (``intrinsic preference'') that requires no access to the model's internal weights or architecture.
We apply Jeffreys smoothing to avoid infinite logits in low-count bins (see Appendix~\ref{app:logit}).
For models with strong bias at low temperature (notably \texttt{mistral:7b}), certain local-field bins become extremely rare; in these cases we supplement the standard binning with a rare-event analysis that pools statistics across neighboring temperature points to stabilize the fit (Appendix~\ref{app:logit}).

\subsection{Simulation parameters}

We study three open-weight models (\texttt{llama3.1:8b}, \texttt{phi4-mini:3.8b}, and \texttt{mistral:7b}) served locally via Ollama~\cite{ollama}.
For finite-size scaling we use even-$L$ lattices ($L=10, 16, 20, 24$ for \texttt{llama3.1:8b} and \texttt{phi4-mini:3.8b}; up to $L=35$ for \texttt{mistral:7b}, even though we use up to $L=30$ to compute $\gamma/ \nu$); the rationale for restricting the analysis to even $L$ is discussed in Appendix~\ref{app:tech}.
Sampler temperatures are scanned from $T=0.05$ to $T\approx 3.2$.
For each $(L,T)$ we run at least six independent seeds, equally split between $g=+1$ and $g=-1$ flip sectors.
The total number of sweeps per run scales as $2L^2$ near the crossover temperature, ensuring that the burn-in window (70\% of total sweeps) comfortably exceeds the system's relaxation time, which grows as $\sim L^2$ for Glauber-like dynamics.
Observables are computed over the remaining 30\% measurement window, and uncertainties are estimated by bootstrap over seeds.
All plots show $\pm 1\sigma$ error bands from the seed-averaging procedure.
All production runs were executed on identical hardware (NVIDIA RTX 5090) to eliminate architecture-dependent variability in LLM inference (see Appendix~\ref{app:tech}).

\section{Results}
\label{sec:results}

\subsection{Magnetization dynamics}

Figure~\ref{fig:mag-timeseries} shows the magnetization time series $m(t)$ for all three models across sampler temperatures.
The qualitative behavior is shared: at low~$T$, the system rapidly reaches high alignment; as $T$ increases, trajectories become noisier and the steady-state magnetization drops.
However, the rate and extent of this disordering differ markedly across models.

\texttt{llama3.1:8b} [Fig.~\ref{fig:mag-timeseries}(a)] reaches near-full alignment ($m\to 1$) at $T\leq 0.2$ and shows a gradual decrease with clear temperature separation.
At intermediate temperatures ($T\approx 0.4$--$0.8$), trajectories are noisy with substantial inter-seed variance, while at $T\geq 1.0$ the magnetization fluctuates around a reduced but still positive value ($m\approx 0.2$--$0.3$), reflecting the residual bias.

\texttt{phi4-mini:3.8b} [Fig.~\ref{fig:mag-timeseries}(b)] shows a similar pattern but compressed in temperature: the transition from high to low magnetization occurs over a narrower $T$ range, consistent with the steep decay of this model's effective field (Sec.~\ref{sec:Jh}).

\texttt{mistral:7b} [Fig.~\ref{fig:mag-timeseries}(c)] is strikingly different.
The system reaches near-complete alignment ($m\to 1$) at \emph{all} sampled temperatures, including $T=3.2$.
This behavior is a direct signature of the persistently strong effective field for this model (Sec.~\ref{sec:Jh}): even at high sampler temperatures, the intrinsic label preference dominates over stochastic noise.

\subsection{Susceptibility and finite-size scaling}
\label{sec:chi-results}

\begin{figure}[t]
\centering
\includegraphics[width=\columnwidth]{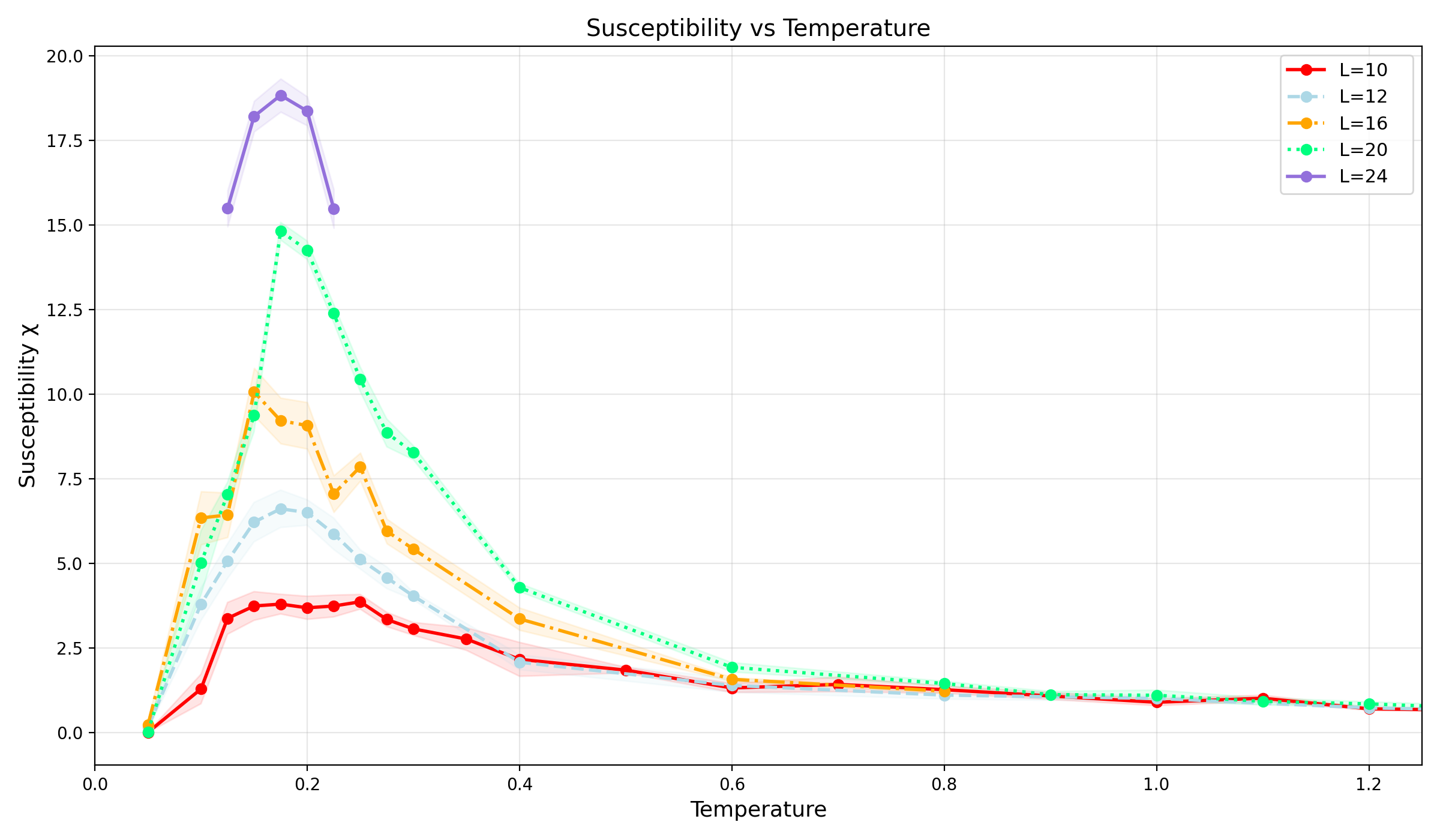}
\caption{Absolute susceptibility $\chi_{|m|}(T)$ for \texttt{phi4-mini:3.8b} on even-$L$ lattices ($L=10,16,20,24$). Shaded bands show $\pm 1\sigma$ seed uncertainty. Peaks are sharp and centered at $T\approx 0.2$, yielding $\gamma/\nu = 1.75 \pm 0.13$. The central value is statistically compatible with the 2D Ising exponent, but the vanishing effective coupling $\tilde{J}\approx 0$ (Fig.~\ref{fig:Jh}) rules out genuine Ising criticality; we interpret the proximity as a coincidence reflecting the steepness of the field crossover (see text).}
\label{fig:chi-phi4}
\end{figure}

\begin{figure}[t]
\centering
\includegraphics[width=\columnwidth]{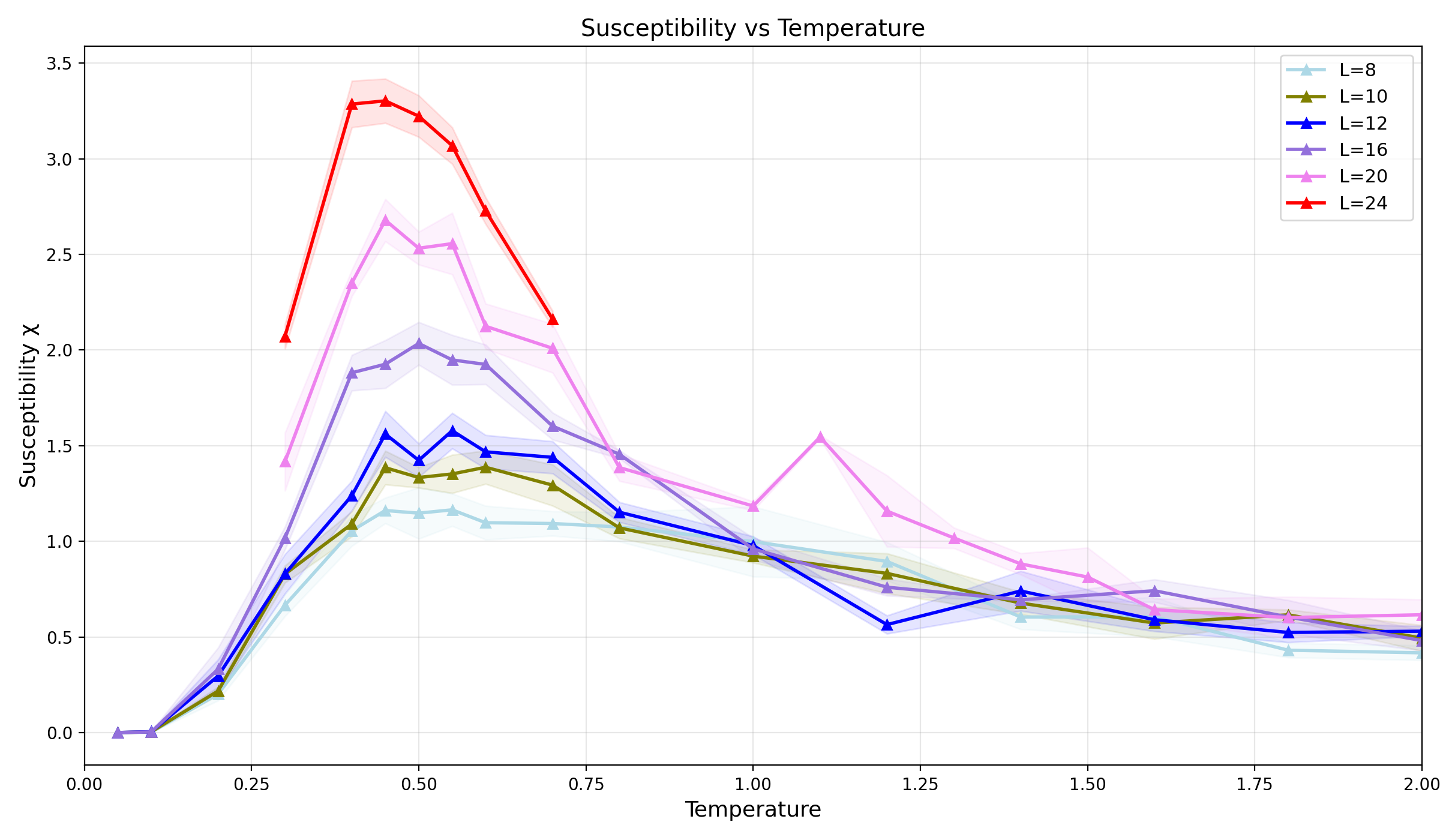}
\caption{Absolute susceptibility $\chi_{|m|}(T)$ for \texttt{llama3.1:8b} on even-$L$ lattices ($L=10,16,20,24$). Shaded bands show $\pm 1\sigma$ seed uncertainty. The effective exponent $\gamma/\nu = 1.02 \pm 0.05$ corresponds to $\chi_{\max}\sim L$, a sub-linear growth indicating weak collective amplification.}
\label{fig:chi-llama}
\end{figure}

\begin{figure}[t]
\centering
\includegraphics[width=\columnwidth]{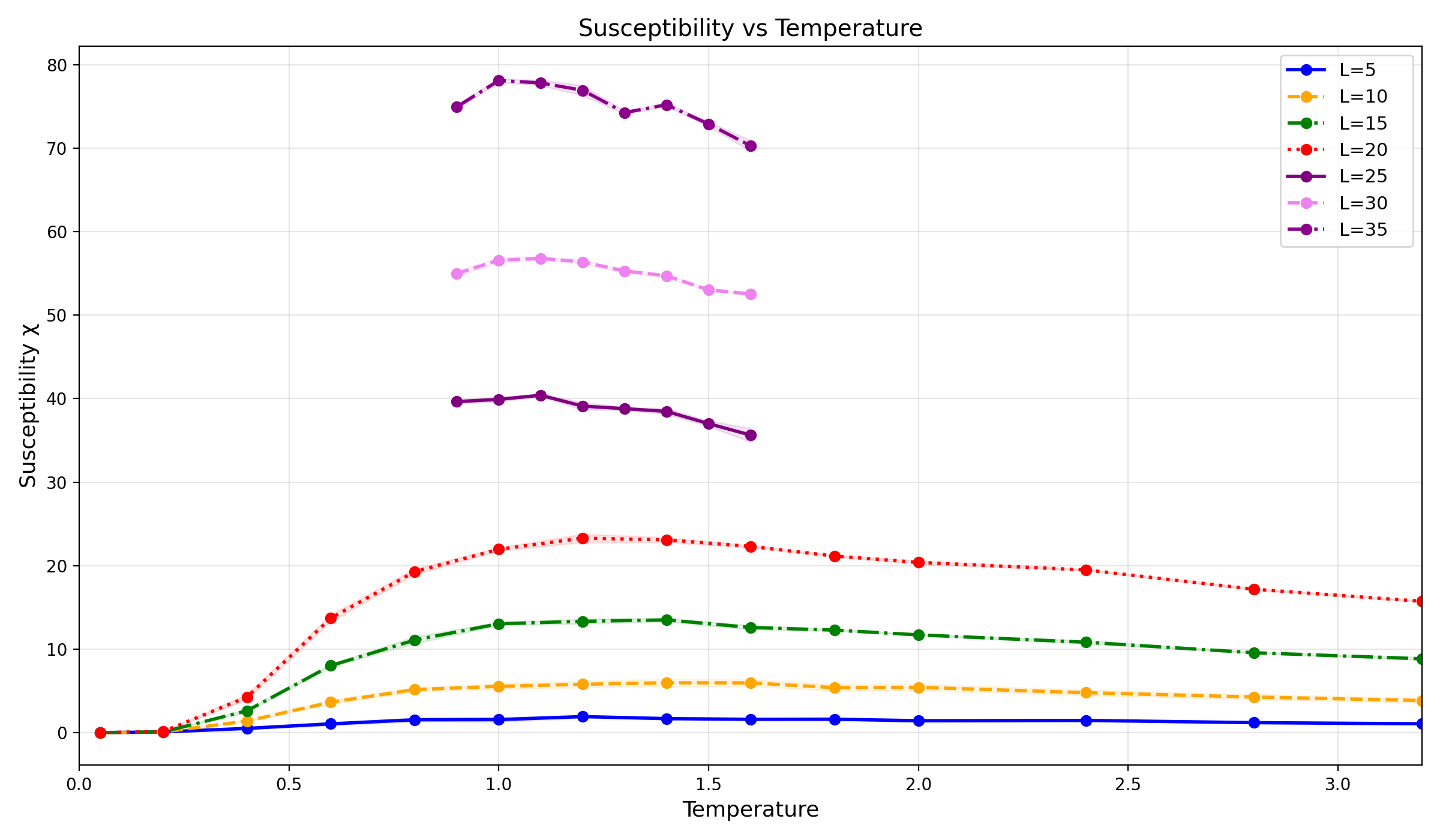}
\caption{Absolute susceptibility $\chi_{|m|}(T)$ for \texttt{mistral:7b} on even-$L$ lattices, up to $L=35$. Peaks are very broad, centered around $T\approx 1.0$--$1.2$, and grow as $\gamma/\nu = 2.01 \pm 0.04$, corresponding to trivial volume scaling $\chi_{\max}\propto N$ ($\gamma/\nu$ are computed on even-L only). This reflects a field-dominated system whose fluctuations scale with size rather than with a diverging correlation length.}
\label{fig:chi-mistral}
\end{figure}

Figures~\ref{fig:chi-phi4}--\ref{fig:chi-mistral} show the absolute susceptibility $\chi_{|m|}(T)$ for all three models on even-$L$ lattices.
Every model exhibits a susceptibility peak, indicating a temperature of maximal consensus fragility.
However, the character of the peaks and the resulting finite-size scaling differ qualitatively, yielding three distinct effective exponents extracted from weighted log-log fits of $\chi_{\max}(L)$:
\begin{align*}
\gamma/\nu &= 1.02 \pm 0.05 \quad \text{(\texttt{llama3.1:8b})}, \\
\gamma/\nu &= 1.75 \pm 0.13 \quad \text{(\texttt{phi4-mini:3.8b})}, \\
\gamma/\nu &= 2.01 \pm 0.04 \quad \text{(\texttt{mistral:7b})}.
\end{align*}
The fact that these exponents are all different is itself strong evidence against universal Ising behavior: in a genuine universality class, the exponent would be the same regardless of microscopic details.

\texttt{phi4-mini:3.8b} (Fig.~\ref{fig:chi-phi4}) displays sharp, well-defined peaks centered at $T\approx 0.2$, with peak height growing cleanly with $L$.
The effective exponent $\gamma/\nu = 1.75 \pm 0.13$ is statistically compatible with the 2D Ising value of $7/4 = 1.75$.
However, this proximity is not evidence for genuine Ising criticality.
As shown in Sec.~\ref{sec:Jh}, $\tilde{J}\approx 0$ for \texttt{phi4-mini:3.8b} at all temperatures, meaning there is essentially no neighbor coupling.
A diverging correlation length, the mechanism that produces $\gamma/\nu = 7/4$ in the Ising model, requires finite coupling.
A system with vanishing $\tilde{J}$ cannot develop the long-range correlated domains characteristic of a critical point.
We therefore interpret the proximity to $7/4$ as a coincidence that reflects the steepness of \texttt{phi4-mini:3.8b}'s field crossover: $\tilde{h}(T)$ drops from $\sim 2.7$ to near zero in a narrow window $\Delta T \approx 0.2$, producing a sharp peak in $\chi_{|m|}$ whose $L$-scaling, over the limited range of accessible system sizes, happens to mimic Ising-like behavior.

\texttt{llama3.1:8b} (Fig.~\ref{fig:chi-llama}) shows intermediate behavior, with peaks around $T\approx 0.4$--$0.5$.
The effective exponent $\gamma/\nu = 1.02 \pm 0.05$ corresponds to $\chi_{\max}\sim L$ (or equivalently $\sqrt{N}$), a sub-linear growth that lies between the regime of independent agents ($\gamma/\nu = 0$) and the volume-scaling regime ($\gamma/\nu = 2$).
Physically, this indicates weak but real cooperative amplification: \texttt{llama3.1:8b} agents are not independent, but their correlations do not extend to the long ranges characteristic of a critical system.

\texttt{mistral:7b} (Fig.~\ref{fig:chi-mistral}) produces the largest susceptibility values ($\chi\sim 80$ at $L=35$) with very broad peaks centered around $T\approx 1.0$--$1.2$ that grow rapidly with system size.
The effective exponent $\gamma/\nu = 2.01 \pm 0.04$ saturates the dimensional ceiling $d=2$ and corresponds exactly to volume scaling $\chi_{\max}\propto L^2 = N$.
This is the expected behavior for a uniformly magnetized system with small Gaussian fluctuations: the variance of the total magnetization scales as $N$ by the central limit theorem, and dividing by $T$ gives $\chi\propto N/T$.
The persistent $\tilde{h}$ for \texttt{mistral:7b} (Sec.~\ref{sec:Jh}) confirms that the system never truly disorders; the broad ``peak'' marks where the field weakens enough for volume-scaled fluctuations to become visible, not a critical point.

\subsection{Effective couplings and fields}
\label{sec:Jh}

\begin{figure}[t]
\centering
\includegraphics[width=\columnwidth]{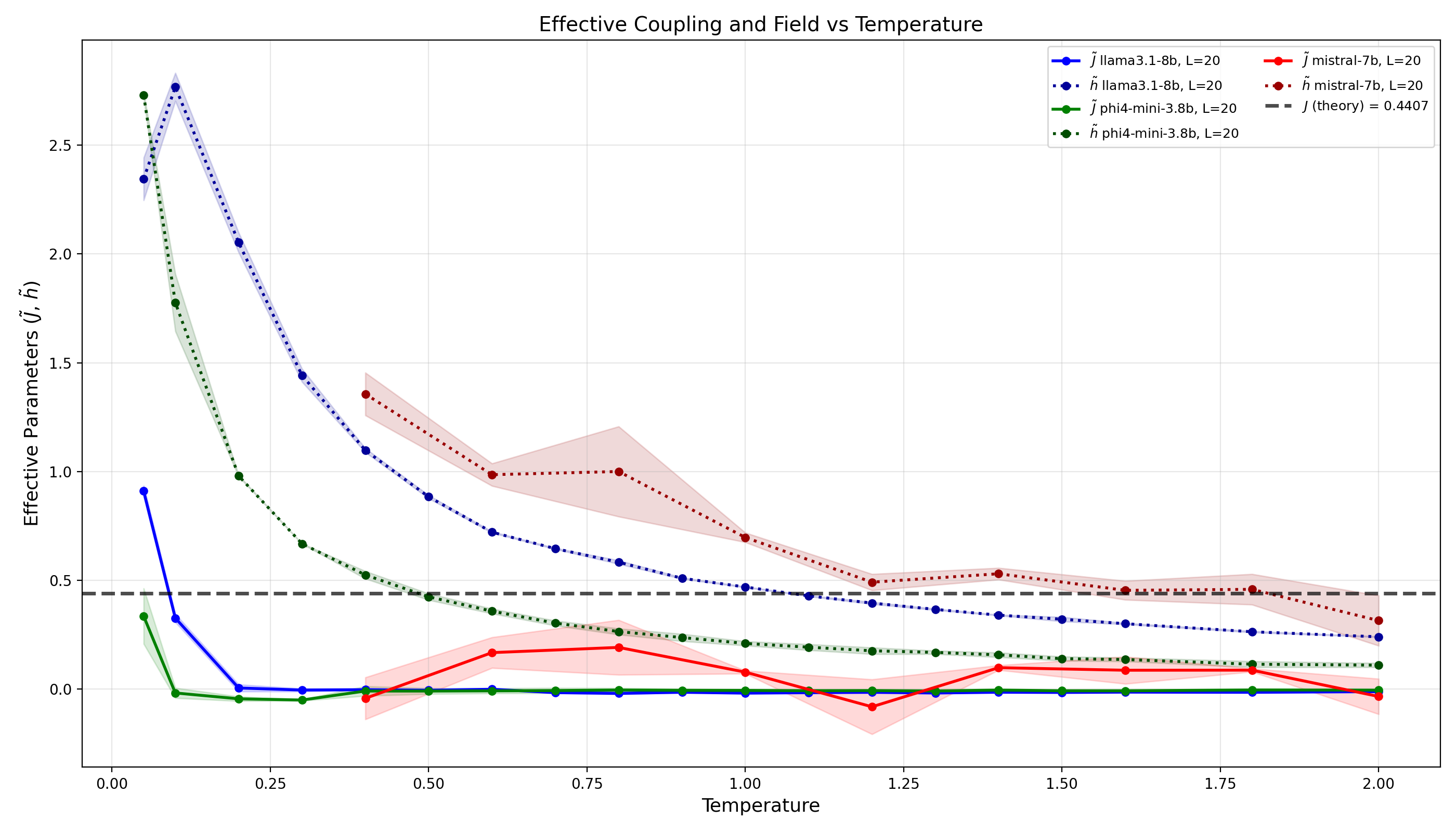}
\caption{Effective coupling $\tilde{J}$ (solid lines) and field $\tilde{h}$ (dotted lines) versus sampler temperature for \texttt{llama3.1:8b} (blue), \texttt{phi4-mini:3.8b} (green), and \texttt{mistral:7b} (red), measured at $L=20$. Shaded bands show $\pm 1\sigma$ seed uncertainty. The horizontal dashed line marks the 2D Ising critical coupling $\beta_c J\approx 0.4407$. For all models, $\tilde{h}\gg\tilde{J}$: collective alignment is field-dominated rather than coupling-driven. The rate at which $\tilde{h}(T)$ decays (steep for \texttt{phi4-mini:3.8b}, moderate for \texttt{llama3.1:8b}, slow for \texttt{mistral:7b}) directly determines the character of the susceptibility peak and the effective $\gamma/\nu$ (Figs.~\ref{fig:chi-phi4}--\ref{fig:chi-mistral}).}
\label{fig:Jh}
\end{figure}

Figure~\ref{fig:Jh} presents the central diagnostic of this study: the effective coupling $\tilde{J}(T)$ and effective field $\tilde{h}(T)$ extracted from update logs for all three models.
The horizontal dashed line marks the 2D Ising critical coupling $\beta_c J \approx 0.4407$.

Across all models and temperatures, $\tilde{h}\gg\tilde{J}$: the field dominates while the coupling remains near zero.
This single observation explains both the absence of Ising criticality and the model-dependent effective exponents reported in Sec.~\ref{sec:chi-results}.
A second-order phase transition in the Ising universality class requires $h= 0$ and finite $J$; when the field dominates, the transition is replaced by a smooth crossover whose steepness depends on the rate at which $\tilde{h}(T)$ decays.

The three models trace qualitatively distinct trajectories in $(\tilde{J},\tilde{h})$ space, and these trajectories map directly onto the susceptibility signatures.
\texttt{phi4-mini:3.8b} (green) exhibits the steepest field decay: $\tilde{h}$ drops from $\sim 2.7$ to near zero by $T\approx 0.3$, with $\tilde{J}\approx 0$ throughout.
This steep crossover produces the sharp susceptibility peak and the exponent $\gamma/\nu = 1.75 \pm 0.13$ of Fig.~\ref{fig:chi-phi4}.
\texttt{llama3.1:8b} (blue) shows intermediate field decay, with $\tilde{J}\approx 0.35$ at the lowest temperature (the only model where the coupling approaches the Ising critical value), but $\tilde{h}\approx 2.8$ at the same point, so the field still dominates by nearly an order of magnitude.
\texttt{mistral:7b} (red) has the slowest field decay: $\tilde{h}$ starts at $\sim 1.3$ (at $T=0.4$) and remains substantial across the full sampled range, while its coupling $\tilde{J}$ fluctuates around $0.1$--$0.2$ with large uncertainty.
This persistent field explains the broad susceptibility plateau and the volume scaling $\gamma/\nu = 2.01 \pm 0.04$ of Fig.~\ref{fig:chi-mistral}. The physical interpretation is sharp: most of the times, what looks like collective consensus at low temperature is not the result of agents influencing each other. It is the result of all agents independently following the same intrinsic bias. This is ``consensus by shared preference,'' not ``consensus by deliberation.''

Taken together, these results rank the three models in terms of genuine social behavior: \texttt{llama3.1:8b} is the most social, with sub-linear scaling reflecting real (if modest) inter-agent influence; \texttt{phi4-mini:3.8b} is essentially asocial, with vanishing $\tilde{J}$
producing an Ising-like exponent that has nothing to do with cooperation; and \texttt{mistral:7b} is the least social, with alignment driven almost entirely by its persistent field.

\section{Discussion}
\label{sec:discussion}

\subsection{Why deviations from Ising universality are expected}

The 2D Ising universality class rests on three conditions: a $\mathbb{Z}_2$-symmetric Hamiltonian ($h=0$), detailed balance (convergence to a Gibbs measure), and short-range interactions.
Our LLM lattice generically violates the first two.

The broken $\mathbb{Z}_2$ symmetry is manifest in the large $\tilde{h}$ values of Fig.~\ref{fig:Jh} and in the asymmetric $P(m)$ distributions of Fig.~\ref{fig:Pm}: the LLM's weights and alignment training encode a systematic label preference that acts as an external field.
The global-flip protocol cancels this bias at the level of averaged even moments, but does not restore microscopic $\mathbb{Z}_2$ symmetry.
Each individual trajectory still evolves under the biased update kernel, so within-run correlations carry the field's imprint.
In equilibrium Ising with $h\neq 0$, there is no phase transition at all; the singularity is replaced by a smooth crossover, which is precisely what we observe.

Detailed balance is the second concern.
The LLM update kernel is an empirical object: while the logistic fit~\eqref{eq:logit} implies consistency with \emph{some} effective Hamiltonian when the fit is good, there is no guarantee that the actual conditional probabilities satisfy detailed balance with respect to any energy function.
If they do not, the stationary distribution (if one exists) is not a Gibbs state, and equilibrium critical exponents need not apply.
Independent evidence for non-equilibrium dynamics in our setup is provided by the lattice-parity effect described in Appendix~\ref{app:tech}: in an equilibrium system, $\chi_{\max}(L)$ is monotonically non-decreasing regardless of parity, whereas persistent probability currents in configuration space can couple to the lattice topology in $L$-dependent ways.

Finally, the LLM's decision is not strictly a function of the scalar local field $k=\sum_{j\in\partial i}s_j$.
The agent processes a token sequence and may be sensitive to the spatial arrangement of neighbors, prompt syntax, and token ordering.
If the effective update rule depends on more than $k$, the system departs from the Ising universality class even when other conditions are met.

\subsection{The sampler temperature as a control parameter}

The sampler temperature rescales logits before the softmax, providing a well-defined noise axis: $T\to 0$ gives deterministic decoding and $T\to\infty$ gives uniform sampling.
However, increasing $T$ simultaneously reduces both $\tilde{J}$ and $\tilde{h}$, at rates that differ across models (Fig.~\ref{fig:Jh}).
The system therefore moves along a \emph{curve} in $(\tilde{J},\tilde{h})$ space rather than along the $h=0$ line where the Ising transition lives.
This explains why each model traces a different crossover trajectory and why no universal $T_c$ emerges.

\subsection{The spectrum of effective exponents}

The three measured exponents ($\gamma/\nu \approx 1.02$ for \texttt{llama3.1:8b}, $\approx 1.75$ for \texttt{phi4-mini:3.8b}, $\approx 2.01$ for \texttt{mistral:7b}) are not random scatter but correlate systematically with the rate at which $\tilde{h}(T)$ decays.
A steep field decay (\texttt{phi4-mini:3.8b}) concentrates the crossover in a narrow $T$ window, producing a sharp susceptibility peak whose $L$-scaling happens to match the Ising value within uncertainty, despite the absence of any genuine critical mechanism.
A slow decay (\texttt{mistral:7b}) keeps the system magnetized at all accessible temperatures, so $\chi_{\max}$ scales trivially with volume ($L^2$).
An intermediate decay (\texttt{llama3.1:8b}) yields a near-linear growth in $L$, reflecting weak collective amplification diluted over a broad crossover.
This pattern suggests a unified interpretation: the effective exponent is not a critical exponent in the renormalization-group sense, but rather a measure of how abruptly the LLM's intrinsic bias switches off as stochasticity increases.

\subsection{Implications for multi-agent systems}

Independent of whether strict Ising universality holds, the $(\tilde{J},\tilde{h})$ decomposition provides actionable diagnostics for LLM multi-agent design.
A large $\tilde{h}/\tilde{J}$ ratio signals that apparent consensus is driven by shared bias rather than genuine collective reasoning, a form of ``echo chamber by default.''
The susceptibility peak locates the temperature of maximal consensus fragility, where small perturbations can most effectively shift group opinion.

The framework also reveals intrinsic properties of the LLM models themselves.
\texttt{mistral:7b} acts as a strongly biased agent that maintains alignment over a wide temperature range, reflecting a model whose token preferences are deeply embedded and resistant to contextual override.
\texttt{phi4-mini:3.8b} transitions sharply from biased order to independent disorder, suggesting a model whose bias is strong but narrowly concentrated in logit space and easily disrupted by temperature.
\texttt{llama3.1:8b} occupies an intermediate regime, showing the weakest but nonzero coupling and the most pronounced non-equilibrium signatures.
These collective-behavior fingerprints expose aspects of each model's decision-making that would be difficult to access through standard single-agent evaluations, and they suggest a new class of benchmarks for comparing LLMs as social agents.

An ideal cooperative agent for multi-agent coordination would have $\tilde{J}>0$ and $\tilde{h}\approx 0$, exhibiting genuine neighbor-responsive behavior without intrinsic bias.
Designing prompts that achieve this balance, while preserving the $\mathbb{Z}_2$ symmetry required by the diagnostic framework, is a concrete target for future prompt engineering.

\subsection{The $\tilde{h}/\tilde{J}$ ratio as a consensus reliability measure}

The ratio $\tilde{h}/\tilde{J}$ has a direct practical interpretation: it measures how much of observed collective alignment is attributable to shared bias versus cooperative interaction.
When $\tilde{h}/\tilde{J}\gg 1$, as in all three models studied here, the multi-agent consensus is effectively a single-agent opinion amplified $N$ times; the collective provides no additional epistemic value over a single query.
Only when $\tilde{h}/\tilde{J}\lesssim 1$ does agreement among agents carry genuine evidential weight, because it reflects inter-agent influence rather than correlated bias.

This diagnostic can be applied before deploying any MAS pipeline.
For instance, \texttt{llama3.1:8b} is the only model in our study with measurable coupling ($\tilde{J}\approx 0.35$ at low $T$), making it the strongest candidate for tasks requiring genuine deliberation, such as iterative debate or collaborative reasoning.
However, even for \texttt{llama3.1:8b}, $\tilde{h}/\tilde{J}\approx 8$ at the lowest temperature, meaning bias still dominates by nearly an order of magnitude.
This sobering finding suggests that, for current open-weight models under minimal prompting, multi-agent consensus should be treated as ``amplified single-agent opinion'' rather than ``deliberated group judgment.''
Achieving the cooperative regime ($\tilde{h}/\tilde{J}\lesssim 1$) likely requires targeted prompt engineering or architectural modifications, and the $(\tilde{J},\tilde{h})$ framework provides the quantitative feedback loop needed to guide such efforts.

\section{Limitations and Outlook}
\label{sec:outlook}

Several limitations constrain our current results.
Lattice sizes ($L\leq 35$) and sweep counts are limited by the cost of LLM inference, restricting finite-size scaling to roughly half a decade in $L$.
Burn-in and equilibration are difficult to certify, and strong prompt dependence means that results may shift with alternative phrasings.
The asynchronous update protocol introduces additional non-equilibrium effects.

These limitations motivate a clear roadmap for future work.
First, prompt design could be optimized to minimize the effective field (``field tuning''), potentially enabling access to the $h\approx 0$ regime where a genuine transition might appear.
Preliminary experiments with modified prompts show that prompt changes can dramatically alter the $(\tilde{J},\tilde{h})$ balance and can break the $\mathbb{Z}_2$ symmetry of the global-flip protocol, highlighting both the promise and the delicacy of this approach.
Second, direct tests of detailed balance (checking whether the empirical transition rates are consistent with \emph{any} energy function) would determine whether an equilibrium description is even approximately valid.
Third, comparisons against known non-equilibrium models (voter dynamics, majority rule) on the same lattice would help classify the dynamics when equilibrium descriptions fail.
Fourth, larger lattices, longer runs, and autocorrelation-time measurements would sharpen finite-size scaling and test whether the near-Ising exponent of \texttt{phi4-mini:3.8b} persists at larger $L$.
Finally, extensions to richer state spaces, longer-range interactions, and heterogeneous agents could connect this framework to more realistic multi-agent coordination scenarios.

\section{Conclusion}
\label{sec:conclusion}

We have introduced a minimal, model-agnostic statistical-physics framework for characterizing collective behavior in 2D lattices of LLM agents with local binary interactions.
The framework requires no access to the model's internal weights or architecture; it operates entirely from observable update logs and is applicable to any LLM that can be queried with a structured prompt.
Through this framework, one can disentangle intrinsic bias from cooperation, compute critical exponents, and more generally explore the statistical and collective behavior, including possible phase transitions, of arbitrary multi-agent systems.

Across three open-weight models, we find robust order-disorder crossovers driven by sampler temperature, susceptibility peaks indicating regions of maximal consensus fragility, and a systematic dominance of the effective field over the effective coupling ($\tilde{h}\gg\tilde{J}$).
This field dominance, reflecting intrinsic label preferences encoded in the LLM's weights, replaces the Ising phase transition with a model-dependent crossover and produces a spectrum of non-universal effective exponents ($\gamma/\nu = 1.02 \pm 0.05$, $1.75 \pm 0.13$, and $2.01 \pm 0.04$ for \texttt{llama3.1:8b}, \texttt{phi4-mini:3.8b}, and \texttt{mistral:7b} respectively) that correlate with the rate at which the effective field decays with temperature.

Rather than a negative result, this is a concrete, quantitative characterization of how LLM collectives reach apparent consensus.
The $(\tilde{J},\tilde{h})$ decomposition reveals that low-temperature alignment is ``consensus by shared preference'' (each agent independently following its bias) rather than ``consensus by deliberation'' (cooperative inter-agent coupling).
The framework also exposes intrinsic properties of the models themselves: \texttt{mistral:7b}'s persistent field reveals deeply embedded token preferences resistant to contextual override, \texttt{phi4-mini:3.8b}'s steep field decay indicates a bias that is strong but easily disrupted by temperature, and \texttt{llama3.1:8b}'s non-equilibrium signatures expose the non-Boltzmann character of its update kernel.
These collective-behavior fingerprints complement standard single-agent evaluations and suggest a new class of benchmarks for comparing LLMs as social agents.

The ratio $\tilde{h}/\tilde{J}$ serves as a direct, practical measure of consensus reliability.
When bias dominates over neighbor-responsiveness, as in all three open-weight models studied here, multi-agent consensus should be treated as amplified single-agent opinion rather than deliberated group judgment.
For multi-agent alignment, the consequence is concerning: architectures that rely on agreement among copies of the same model (debate, self-consistency, committee voting) cannot assume that $N$-agent consensus is more informative than a single query when $\tilde{h}\gg\tilde{J}$, because the agreement is correlated through shared bias rather than produced by independent reasoning.
This concern is especially acute in decentralized multi-agent systems, where no central authority can correct for shared biases, and where the emergence of apparent consensus may be mistaken for robust collective intelligence.

The susceptibility peak provides a complementary diagnostic: it identifies the operating temperature at which the collective is most responsive to external steering, offering a principled way to design human-in-the-loop protocols that can most effectively guide group opinion.
Together, the $(\tilde{J},\tilde{h})$ decomposition and the susceptibility analysis form a compact toolkit for evaluating, comparing, and improving LLM multi-agent systems.

This work establishes a starting point for a statistical-physics and condensed-matter inspired research program on LLM multi-agent systems, and motivates targeted investigations into prompt-based field tuning to reach the genuinely cooperative regime ($\tilde{h}/\tilde{J}\lesssim 1$), collective decentralized alignement, detailed-balance diagnostics to determine when equilibrium descriptions apply, and non-equilibrium universality in collective artificial intelligence.

\appendix

\section{Global-flip mapping details}
\label{app:flip}

For each simulation run, a global parity $g\in\{-1,+1\}$ is drawn. When $g = +1$ nothing happens. When $g=-1$, all prompts swap \texttt{yes}$\leftrightarrow$\texttt{no} in both the neighbor-state description and the response parsing: an LLM reply of \texttt{yes} is recorded as $s_i=-1$, and vice versa.
Observables that are even in $m$ ($\langle m^2\rangle$, $\langle m^4\rangle$) are invariant under this relabeling by construction.
Averaging odd quantities ($\langle m\rangle$, $\langle |m|\rangle$) across $g=\pm 1$ sectors reveals the magnitude of the effective field: a large difference $\langle m\rangle_{g=+1} - \langle m\rangle_{g=-1}$ signals strong label bias.

\section{Logistic regression and rare-event analysis details}
\label{app:logit}

For each $(T, \text{model})$, we aggregate all micro-updates across seeds and lattice sweeps.
Updates are binned by the local field $k\in\{-4,-2,0,2,4\}$, and the empirical probability $\hat{p}_+(k)$ is computed with Jeffreys smoothing: $\hat{p}_+(k) = (n_+(k)+\tfrac{1}{2})/(n(k)+1)$, where $n_+(k)$ is the count of $s_i'=+1$ outcomes and $n(k)$ the total count at field value $k$.
A weighted linear regression of $\operatorname{logit}\,\hat{p}_+(k)$ on $k$ yields $\tilde{h}$ (intercept divided by 2) and $\tilde{J}$ (slope divided by 2).
Weights are proportional to $n(k)$, ensuring that well-sampled field values dominate the fit.
Bins with $n(k)<5$ are excluded as a robustness check; results are stable to this threshold.

For \texttt{mistral:7b}, the strong effective field at low and intermediate temperatures means that local-field bins opposing the preferred label (e.g., $k=-4$ when the bias favors $+1$) are populated by very few updates.
Standard binning yields unreliable logit estimates in these rare-event bins.
We address this by pooling update counts across a sliding window of neighboring temperature points, providing sufficient statistics for stable logistic regression while preserving the temperature dependence of $\tilde{J}$ and $\tilde{h}$.
The larger error bands for \texttt{mistral:7b} in Fig.~\ref{fig:Jh} reflect the residual uncertainty from this procedure.

\section{Technical details and reproducibility}
\label{app:tech}

\paragraph{Even-$L$ lattices and the parity effect.}
All susceptibility scaling analyses presented in this work use even-$L$ lattices ($L=10, 16, 20, 24$ for \texttt{phi4-mini:3.8b} and \texttt{llama3.1:8b}; up to $L=30$ for \texttt{mistral:7b}).
This choice is motivated by the fact that even-$L$ lattices with periodic boundary conditions are bipartite, which preserves the antiferromagnetic symmetry of the underlying Hamiltonian and produces cleaner finite-size scaling.
Odd-$L$ lattices, by contrast, contain frustrated loops induced by the periodic boundary conditions and can introduce subleading topological corrections that distort scaling fits at small system sizes.
In our exploratory runs we observed a systematic lattice-parity effect: odd-$L$ and even-$L$ subsequences yielded different effective exponents, with the largest discrepancy observed for \texttt{llama3.1:8b}.
We interpret this parity effect as a finite-size artifact amplified by the non-equilibrium character of the LLM update kernel: in an equilibrium system, $\chi_{\max}(L)$ is monotonically non-decreasing regardless of parity, but in a non-equilibrium system the persistent probability currents in configuration space can couple to the lattice topology in $L$-dependent ways.
We report only the even-$L$ results in the main text.

\paragraph{Seeds and uncertainty estimation.}
For each $(L, T)$ point, observables are averaged over at least six independent seeds, equally split between the $g=+1$ and $g=-1$ flip sectors.
Uncertainties on $\chi_{\max}(L)$ are estimated by bootstrap resampling over seeds.
The effective exponent $\gamma/\nu$ and its uncertainty are extracted from a weighted least-squares fit of $\log \chi_{\max}$ versus $\log L$, with weights $w_i = 1/\sigma_i^2$, where $\sigma_i = \sigma_{\chi,i}/\chi_{\max,i}$ is the propagated $1\sigma$ uncertainty on $\log \chi_{\max}(L_i)$.
The reported uncertainty on $\gamma/\nu$ is the standard error of the slope returned by this weighted fit.

\paragraph{Hardware reproducibility.}
All production runs were executed on identical hardware (NVIDIA RTX 5090 GPUs) to eliminate architecture-dependent variability in LLM inference.
We observed during preliminary testing that runs of the same model with identical seeds and prompts can yield slightly different logits when executed on different GPU architectures, due to non-associativity of floating-point arithmetic and architecture-specific kernel implementations.
While these differences are negligible for most LLM applications, they can compound over the many micro-updates in our simulations and produce small but systematic shifts in sensitive quantities such as susceptibility and its peaks $\chi_{\max}$.
Restricting all production runs to identical hardware ensures that finite-size scaling comparisons across different $L$ values are not contaminated by this source of variability.

\paragraph{LLM inference calls.}
Across all three models, the full study comprised approximately $10^9$ individual LLM inference calls.
For each $(L, T, \text{seed})$ run, the number of calls is $N \times 2L^2 = 2L^4$
(one call per site per micro-update, with $N=L^2$ sites and $2L^2$ sweeps),
so each lattice size contributes $240 L^4$ calls when averaged over 6 seeds
and scanned over 20 temperatures.
This yields $\sim 1.4\times 10^8$ calls for each of \texttt{llama3.1:8b}
and \texttt{phi4-mini:3.8b} (lattice sizes $L \in \{10, 12, 16, 20, 24\}$),
and $\sim 7.0\times 10^8$ calls for \texttt{mistral:7b} (lattice sizes
$L \in \{10, 15, 20, 25, 30, 35\}$). This is the count for production runs only. If you ran additional exploratory runs, the actual total is somewhat higher, probably closer to $3\times10^9$. This computational cost is the primary reason for restricting finite-size
scaling to roughly half a decade in $L$.

\bibliographystyle{apsrev4-2}
\bibliography{refs}

\end{document}